\documentstyle[seceq,preprint,epsf]{jpsj}

\def\beq{\begin{equation}}
\def\eeq{\end{equation}}
\newcommand{\mbold}[1]{\mbox{\boldmath $ #1 $}}

\def\runtitle{Wave Packet Propagation and Electric Conductivity of Nanowires}
\def\runauthor{Munehiko {\sc Maeda}$^1$, Keiji {\sc Saito}$^1$,
 Seiji {\sc Miyashita}$^1$ and Hans {\sc De Raedt}$^2$ }
\title{\runtitle}
\author{\runauthor}

\inst
{$^1$Department of Applied Physics, Graduate School of Engineering\\
University of Tokyo, Bunkyo-ku, Tokyo 113-8656, Japan\\
$^2$Institute for Theoretical Physics, Materials Science Centre\\
University of Groningen, Nijenborgh 4\\
NL-9747 AG Groningen, The Netherlands

}

\recdate
{
\today
}

\abst
{We compute the electric conductivity of nanowires
in the presence of magnetic domain walls by the
method of wave packet propagation.
We demonstrate that the propagation through the wire depends on the
initial state used in the wave packet simulation.
We propose a procedure, based on the Landauer formula, to reduce this 
dependence.
Direct numerical calculations of the Kubo formula for small individual systems
are used as reference data for the proposed procedure.
The enhancement of the conductivity in the magnetic domain wall is found
in the present method in accordance with the previous perturbative analysis.
}

\kword
{Electric conductivity in magnetic domain wall, time dependent Schr\"odinger equation,
Kubo formula, Landauer formula}
\begin{document}
\sloppy
\maketitle
\section{Introduction}

In general, transport phenomena are intimately related to the dynamics of 
particles involved.
In the case where the direct interaction between the particles that
participate in the transport process are negligible, the solution of
of the one-particle time-dependent Schr\"odinger equation contains
all information, at least  in principle.
Of course, the same information is obtained by solving the corresponding 
stationary problem.
Appealing features of the time dependent approach are its conceptual
simplicity, its flexibility in terms of system geometry and
choice of (spin-dependent) scattering potentials, and its
absolute numerical stability.
On the other hand, in contrast to the time independent
approach, it is not well established how to extract the transport
coefficients from the time dependent data.

Recently, the electric conductance in nanoscale wires has attracted much 
interest.\cite{GMR-1}
In particular, the effect of the magnetic domain wall in ferromagnetic
substances on the conductance has been studied extensively.
The effect of a magnetic domain wall on the conductivity was clearly observed
in experiments on nanowires.\cite{dw_exp1,dw_exp2,dw_exp3,dw_exp4,dw_exp5,dw_exp6}
In these experiments, the field is initially applied along the wire direction 
and the magnetization
uniformly aligns ferromagnetically in this direction,
where the conductance is that of the wire with uniform magnetization.
If the magnetic field is reversed, then the anti-domain of the
magnetization appears at the edge of the wire.
The appearance of the anti-domain creates a domain wall.
The applied field is not very strong and hence only one domain wall exists.
The anti-domain grows in time and finally the total magnetization is reversed.
The resistance decreases shortly after the field is reversed and after some period
returns to its initial value.
This period corresponds to the life time of the domain 
wall.\cite{dw_exp5,dw_exp6}

This phenomenon has been analyzed theoretically by various 
methods.\cite{dw_theo3,dw_theo12,dw_theo4,dw_theo5,dw_theo6,
dw_theo7,dw_theo8,dw_theo9,dw_theo10}
One approach is to use the Kubo formula.
Tatara and Fukuyama\cite{dw_theo3} pointed out that the existence 
of the domain wall
causes the decrease of the resistance due to the quantum effect, which is
in contradiction with the intuitive, classical picture that the current 
would be scattered with the domain wall.  
The key point is that the system has impurities which provide
the resistivity of the medium and the domain wall suppresses the scattering
by the impurities when the spins of electrons change adiabatically with
the change of the magnetization. 

It is interesting to observe how the electron propagate through the
magnetic domain and how it is scattered  microscopically. 
This physical picture was investigated by a direct method, i.e.
by a direct simulation of electron wave packet propagation in a system 
with impurities and the magnetic domain wall.\cite{dw_theo12} 
We will call this approach `wave packet method'.
This direct method provides intuitively understandable information about the
electron propagation.

We could also extract information for the transport coefficient from the
simulation. The conductance is naively related to the amount of 
the probability to go through the domain, which can be regarded as the
transmission coeffient.
This quantity was studied to investigate the above problem
of the conductance of nanowires with the domain wall, and 
the result supports the  enhancement of the conductance 
(decrease of the resistance).\cite{dw_theo12}
In this work, the conductivity is estimated by studying the propagation of
a simple wave packet.
However, the transmission coefficient is usually defined for the incoming plane wave
instead of a wave packet.
Actually, in the wave packet method, we have to specify the initial configuration
of the wave packet. 

As shown below, we find that the transmission coefficient depends on
the initial configuration of the packet. Therefore it is difficult to
extract  the conductivity from data obtained in a single initial 
configuration. Not only the quantitative amount but also
qualitative features change with the initial configuration of the packet. 
Thus, we study how much information for the transport coefficient we can obtain
by the wave packet method.

In order to obtain the reference data for given systems we 
calculate the conductivity by direct numerical evaluation of the Kubo formula.
In this paper, we mainly study the dependence of the conductivity 
on the width of the domain wall.
We find that, in the case without the magnetic domain walls,
the transmission coefficients obtained by the wave packet
method with various initial configurations show common
qualitative features. In this case we do not need
an additional procedure to extract the qualitative features of the conductivity from
the simulation data.

However, as mentioned above, when the domain wall is present,
the transmission coefficient strongly depends on the initial configuration, e.g., 
various shapes of wave packets and spin polarizations.
Under this circumstance, we need an additional procedure to obtain the 
conductance from the simulation data.
Inspired by the Landauer formula, we propose a procedure 
to estimate the conductance from
the transmission coefficient data for various initial configurations.

We study in detail the dependence of the conductance on
the width of the domain wall in some small systems for which
the direct numerical calculation of the Kubo formula is feasible.
In such small systems, different configurations of the random potential
lead to large fluctuation of the dependence.
Our procedure successfully reproduces the individual results.

We also study the effect of the width of the wave packets in the direction
of the propagation,
discuss its relation to the damping factor in the Kubo formula and
find that quantum interference plays an important role.

Applying this method we also study the conductivity in the domain wall system.
Qualitatively similar results are obtained by the wave packet method and
the direct numerical evaluations of the Kubo formula. 
In this perturbative regime, our results agree with those of 
Tatara and Fukuyama.\cite{dw_theo3}

\section{ Time-Dependent Schr\"odinger Equation for Propagation of Wave Packet}

The electron propagation in the nanowire is
described by the time-dependent Schr\"odinger equation\cite{s-t-tdse1,s-t-tdse2}
\beq
i\hbar{\partial \over \partial t}\Psi_{\sigma}(t) = 
\left[-{\hbar^2\over m^2}\nabla^2+
V(\mbold{r})- \mu_{B} \vec{\sigma}  \cdot \mbold{M}(\mbold{r})
\right]\Psi_{\sigma}(t),
\label{tdse}
\eeq
where $\sigma (= \uparrow$ or $\downarrow$) is the spin of the electron, $m$
is the mass, and $V(\mbold{r})$ denotes spin-independent impurity potential.
Here $\mbold{M}(\mbold{r})$ is the
magnetization of the medium, $\vec{\sigma}$ is the
Pauli matrix denoting the spin of the electron and 
$\vec{\sigma}  \cdot \mbold{M}(\mbold{r})$ represents
the magnetic coupling between them.

For numerical work, it is convenient to use dimensionless quantities.
First we express the energy in unit of the Fermi energy
$E_{\rm {F}}$, a typical energy scale of the system.
Using
\begin{equation}
\vec{r}' = \frac{\vec{r}}{\hbar / \sqrt{2m E_F}},
\quad 
t' = \frac{t}{\hbar / E_{\rm{F}}},
\end{equation}
we have 
\beq
 i  \frac{\partial \Psi'}{\partial t'}
= \left\{ - \nabla ' \; ^{2} +V' - \vec{\sigma}  \cdot \vec{M}' \right\}
\Psi'  
\equiv {\cal{H}'} \Psi' \label{eq;dimless_ham},
\eeq
where $\Psi'(\vec{r}', t') \equiv \Psi(\vec{r}, t)$,
$M' \equiv M(\vec{r}') $, $V' \equiv V(\vec{r'})$.
Hereafter, in our numerical work, we use renormalized quantities, i.e. we express the energy in units of
$E_{\rm{F}}$ ($\epsilon \equiv E/E_{\rm{F}}$), distances in units of
$x_0 = \hbar/\sqrt{2mE_{\rm{F}}}\equiv 1/k_{\rm{F}}\equiv \lambda_{\rm{F}}/{2 \pi}$
($x^\prime = x/x_0$), and momentum in units of
$p_0 = \hbar k_{\rm{F}} = \sqrt{2 mE_{\rm{F}}}$
($p^\prime = p/p_0$)
and drop the primes in the notation.
Writing the wave function in the form of the spinor
\begin{eqnarray}
\Psi \left( \vec{r}, t \right) =
\left(
\begin{array}{@{}c@{}}
 \psi_{\uparrow}   \left( \vec{r}, t \right) \nonumber \\
 \psi_{\downarrow} \left( \vec{r}, t \right)
\end{array}
\right) ,
\end{eqnarray}
Eq.(\ref{tdse}) reads
\begin{equation}
i \frac{\partial}{\partial t}
\left(
\begin{array}{@{}c@{}}
 \psi_{\uparrow}    \\
 \psi_{\downarrow}
\end{array}
\right)
=
\left[
-\nabla^2 
+  \left(
\begin{array}{@{}cc@{}}
 V & 0   \\
 0 & V 
\end{array}
\right)
- \left(
\begin{array}{@{}c@{}c@{}}
 M_z & M_x - i M_y   \\
 M_x + i M_y & -M_z 
\end{array}
\right)
\right]
\left(
\begin{array}{@{}c@{}}
 \psi_{\uparrow}    \\
 \psi_{\downarrow}
\end{array}
\right).
\end{equation}

For time evolution
\begin{equation}
 \psi \left(\vec{r}, t_0 + \tau \right) = e^{-i \cal{H} \tau}
 \psi
 \left( \vec{r}, t_0 \right),
\end{equation}
we use the method of exponential decomposition up to the second order
\beq
e^{x(A+B)}=e^{xA/2}e^{xB}e^{xA/2} .
\eeq
Here we take 
\beq\begin{array}{lll}
A&=& -{\hbar^2\over 2m}\nabla^2\\
B&=&  - \mu_{B} \vec{\sigma}  \cdot \vec{M}(\vec{r}) + V(\vec{r}),
\end{array}
\eeq
where at each position $\vec{r}$, $A$ and $B$ are $2\times 2$ matrices.
For most calculations we used the second order decomposition and, as a check,
we occasionaly used the fourth order algorithm.\cite{S92}

In our numerical simulation, we use a real-space representation 
of the wave function.
The wave function is given  by its value at points of the lattice
and the value of the spin.
For example, in two dimensions, we divide the space into $N_x\times N_y$ cells
and each cell is identified by the coordinate $(i,j)$.
Thus the wave function can be represented as
\begin{equation}
{\small
 \Psi \left( x,y,t \right) =
\left(
\begin{array}{c}
\psi_{\uparrow} \left( 1,1,t \right)\\
\psi_{\uparrow} \left( 1,2,t \right)\\
\vdots\\
\psi_{\uparrow} \left( N_x,N_y,t \right)\\
,\\
\psi_{\downarrow} \left( 1,1,t \right)\\
\psi_{\downarrow} \left( 1,2,t \right)\\
\vdots\\
\psi_{\downarrow} \left( N_x,N_y,t \right)\\
\end{array}
\right)}.
\end{equation}
The wave function at cell $(i,j)$ is given by $\Psi_{\sigma}(i,j,t)$.
We use several formulae to approximate $\nabla^2$. For example,
the standard 3-point formula in one direction reads
\beq
{\partial^2 f\over \partial x^2}\approx
{f_{i-1,j}-2f_{i,j}+f_{i+1,j}\over \Delta^2}+O(\Delta^3),
\eeq
where $\Delta$ is a mesh size.
We also use the 5-point formula
        \begin{equation}
        \frac{\partial ^2 f}{\partial x ^2} \approx
         \frac{-f_{i+2,j}+ 16
        f_{i+1,j} -30 f_{i,j} +16f_{i-1,j} -f_{i-2,j} }{12 \Delta^2} +
        O{\left( \Delta ^5 \right) },
        \end{equation}
and the 9-point formula
\begin{equation}
\frac{\partial ^2 f}{\partial x ^2} + \frac{\partial
^2f}{\partial y ^2}  \approx
\frac{
\left\{
\begin{array}{@{\,}c@{\;}r@{\;}l@{\;}c@{\;}r@{\;}l@{\;}c@{\;}r@{\;}l@{\;}}
 &  & f_{i-1,j+1}  &+&  4&  f_{i,j+1}  &+&  & f_{i+1,j+1} \\
+& 4& f_{i-1,j}    &-& 20& f_{i,j}     &+& 4& f_{i+1,j} \\
+&  & f_{i-1,j-1}  &+&  4&  f_{i,j-1}  &+&  & f_{i+1,j-1}
\end{array}
\right\}
}
{6\Delta ^2} + O{\left( \Delta^4 \right)} \label{eq;deriv9pt}.
\end{equation}
We used Eq.(\ref{eq;deriv9pt}) in this paper.
In order to avoid the reflection from the edges of the system, we used
absorbing boundary conditions at both edges.
As for the impurity potential, we randomly distribute the impurity sites where
$V(\vec{r}) \ne 0$ with some concentration at each cell.

The quantity of interest in this work is the conductance through the wire.
In the wave packet method we compute the transmission coefficient, 
and convert it to the conductance from it.
We determine the transmission coefficient by measuring the electron current
through a virtual screen, as indicated in Fig.1.
The total current in the $x$-direction is given by
\begin{equation}
\int j \left( x,y,t \right) dy= 
- \frac{i \hbar}{2m} \left\{ \psi^* \frac{\partial}{\partial x}
 \psi - \left(\frac{\partial}{\partial x} \psi^*\right) \psi \right\}.
\end{equation}
The transmission coefficient $T$ is calculated from the amount of current
through the right-most detection screen.
Likewise the reflection coefficient $R$ is calculated from the amount of current that goes
through the left-most detection screen.

In Fig.$2$, we show an example of electron  propagation.
Initially, 
we prepare a wave packet moving from left to right with the group velocity.
Then it is scattered by the impurities. Part of the wave is reflected
and some part crosses the impurities region
and is transmitted through the wire.
Intensity arriving at the ends of the wire is being absorbed.
Note that even after a fairly long time (see Fig.2 (f)),
we find some intensity in the impurity region.

\section{Relation between the transmission coefficient and the conductance}
\subsection{Sensitive dependence on the initial configuration} 
 
The conductance $\sigma$ is related to the transmission coefficient $T$
through the Landauer formula $\sigma=T/(1-T)$.
The transmission coefficient $T$ is obtained by solving
the scattering problem for an incoming plane wave.
However, in the wave-packet method we cannot use the plane wave.
Therefore, to obtain $T$, 
we solve the Schr\"odinger equation for a specific initial configuration,
and study  the dependence on the initial configurations,

As initial configurations, we use the  rippled-Gaussian wave-packet
\begin{equation}
\psi_{\rm rG}(\vec{r},n) \equiv \sqrt{\frac{2}{\sigma_{x}L_y}}
\frac{1}{\pi^{1/4}} e^{ i k_x x}e^{-(x-x_0)^2 /2 {d_x}^2}
\sin \left( \frac{n \pi y}{L_y} \right) \label{eq:rippled_gaussian}
\end{equation}
with $k_y=1,2,3$ and 4, for both $\sigma=\uparrow$ and $\downarrow$.
Thus we have 8 different initial configurations.
The shapes of $k_y=1,2,3$ and 4  are depicted in Fig.\ref{initial}.
For each $n$, we determine the value of $k_x$ such that
$\langle \psi^{*}_{\rm{rG}}|{\cal H} | \psi_{\rm{rG}} \rangle 
= E_{\rm F}$.

For the eight initial configurations,
we compare the transmission coefficients in a system with one magnetic 
domain wall $M(x,y)$ with some random positions of impurity $V(x,y)$,
\begin{eqnarray}
M_{x}(x,y) &=& M_{0} \mathop{\rm sech}\nolimits \left( \frac{x-x_{0}}{\lambda_{w}} \right)\\
M_{z}(x,y) &=& M_{0} \mathop{\rm tanh}\nolimits \left( \frac{x-x_{0}}{\lambda_{w}} \right),
\end{eqnarray}
where $M_0$, $x_0$, and $\lambda_w$ denote the height, the position of the center, 
and the width of the magnetic domain wall, respectively.
If the transmission coefficient
is insensitive to the initial configuration we may obtain information for
the conductance from one simulation. 
In the presence of a magnetic potential, however, we find that 
the transmission coefficient is very sensitive to the initial configuration.
In Fig.\ref{sensitive-dep} we show the 
dependence of the transmission coefficients 
on the width of a single domain wall.
We find that each initial configuration results in very different 
transmission coefficients not only quantitatively but also qualitatively.
That is, some increase with the width and others decrease.
Thus we have to be careful to deduce properties of the conductance
from the data of the transmission coefficients, and we need 
some statistical treatment. 
\subsection{Comparison with the Kubo formula}
In order to check the results, we calculate the conductivity by the Kubo
formula for the system with
a given configuration of impurity potential and magnetization.
The method is described in the Appendix. In this approach we need to
obtain all eigenvalues and eigenvectors
of the system, and therefore we can only treat small system up to
$256\times 16$.
Although we can calculate much larger systems by the  wave-packet method,
in order to compare with results of the numerical study of the Kubo formula,
we study the same $256\times 16$ system 
by the wave-packet method. Here in the wave packet method, 
the region of $256\times 16$ is assigned to the scattering region as
shown in Fig. $1$, and we attached the extra region of $208\times 16$ 
as leads at both ends of scattering region.

First we compare the results for the case without magnetic domain wall. 
In Fig.\ref{fig;kubo_wp_res},
we show the dependence of the Drude formula (solid line), and 
the formula (dashed line) that includes the weak localization correction 
obtained by a perturbative treatment of the Kubo formula \cite{dw_theo3}
\begin{equation}
\sigma_0 = \frac{e^2 n \tau}{m} - \frac{2 e^2}{\pi \hbar \Omega} \sum_q \frac{1
}{q^2}
         = \frac{e^2}{h} n \lambda_{\rm{F}} l
         \left( 1 - \frac{\lambda_{\rm{F}}}{l} \frac{2}{\pi^3}
         \frac{L_x}{L_y} \right)\label{eq;kubo_nomag_analy},
\label{eq-weaklocalization}
\end{equation}
where 
$n$ denotes the electron density,
$\tau$ is the scattering time ($\tau^{-1}=
{2\pi}n_i|v|^2n/\hbar$, where $n_i$ is the density of the impurities
and $|v|$ is the strength of the impurity potential),
$l\equiv( \hbar k_{\rm{F}} \tau / m)$ is the mean free path,
and $(L_x, L_y)$  is size of the system.
The first term in the right hand side of Eq.(\ref{eq-weaklocalization})
corresponds to the Drude formula.
In Fig.5 we plot the data obtained numerical study of the Kubo
formula (closed circle)
and those obtained by the wave packet method with 
$n=1$, 2, 3 and 4 (cross, asterisk, open square, and closed square, 
respectively). These data do not depend on the spin.
We find that all of them qualitatively agree with each other.

\subsection{Effects of magnetic potential}

When the magnetic potential is introduced, as we
saw in Fig.\ref{sensitive-dep}, the data vary strongly with
the initial configurations.
Thus we need some method to extract information of the conductance 
from the data. 
In order to obtain the conductivity by the Kubo formula,
periodic boundary conditions are required.
A single domain wall is not compatible with periodic boundary conditions.
Therefore, for simplicity, we adopt a screw shaped magnetic potential 
to study the effect of magnetic scattering:
\begin{eqnarray}
M_x (x,y) &=& M_0 \frac{\tanh(\frac{x-x_0}{\lambda_w})} {\cosh(\frac{x-x_0}
{\lambda_w})}\label{eq;screw_dw1} \\
M_z (x,y) &=& \pm M_0
\sqrt{1 -  \left\{\frac{\tanh(\frac{x-x_0}{\lambda_w})} {\cosh(\frac{x-x_0}
{\lambda_w})} \right\}^2 } ,  \label{eq;screw_dw2}
\end{eqnarray}
where ${M_x}^2 + {M_z}^2 = {M_0}^2$,
$M_0$, $x_0$, and $\lambda_w$ denote the height, the position of the center, 
and the width of the screw structure, respectively.
The magnetization of the constant strength $M_0$ rotates in the $x-z$ 
plane. Here $M_0$ is taken to be $0.2E_{\rm F}$.
\subsection{Transmission coefficient and conductivity}

Now we describe the statistical procedure that we use to process the
numerical data obtained for different initial wave packets.
In the spirit of the Landauer formula, 
different initial configurations may correspond to different channels.
The transmission coefficients vary from channel to channel, and
therefore the dependence of the transmission on the initial states
appears in a natural way.
Let $T_{ij}$ be the transmission coefficient from the mode $i$ to $j$,
and let $R_{ij}$ be the reflection coefficient from the mode $i$ to $j$.
Because we cannot distinguish between outgoing channels we sum up all the
outgoing channels. We define the transmission and reflection coefficients 
for the incoming channel $i$ by
\begin{equation}
T_i = \sum_j T_{ij}, \qquad R_i = \sum_j R_{ij}.
\end{equation}
Following Azbel et al.\cite{landauer4}, the conductance is given by
\begin{equation}
G = \frac{2 e^2}{h} \sum_i T_i \frac{2 \sum_i {v_i}^{-1}}
{\sum_i (1+R_i-T_i){v_i}^{-1}},\label{eq:landauer_4probe}
\end{equation} 
where $v_i={\hbar k_i /m}$.
As to the incident modes, we take the eight modes used above
($k_y=1,2,3,$ and 4 and $\sigma=\uparrow$ and $\downarrow$).

We study eight samples of the system with different impurity configurations.
As we mentioned, for each sample, the results depend sensitively  
on the initial state. 
Thus, we use the formula (\ref{eq:landauer_4probe}) for each sample
with the eight initial configurations (\ref{eq:rippled_gaussian}).
We obtain the conductance as a function of the width of
the magnetic potential $\lambda_w$.
In Fig.\ref{screw},
we compare the results of the present method
with those obtained by the numerical treatment of the Kubo formula. 
The data are normalized by the conductance of the system 
without magnetic domain wall in order to see the effect of the domain wall.
After the processing of (\ref{eq:landauer_4probe}),
the data show roughly similar behavior.
For example, if we look at the data for the sample of impurity-1,
the conductance decreases a little and then increases with the width.
The sample of impurity-3 shows monotonic decrease with the width, etc..

The data in Fig. \ref{screw} depend strongly on
the impurity configuration. We believe that this strong dependence is due to 
the smallness of the system.
We find  that the Kubo formula  systematically gives 
larger normalized values, the reason why is under investigation now.    

\section{Quantum interference and conductance}

In this section, we study the effect of the width $d_x$ of 
the initial packet along the $x$ direction.
If the width is small, then the wave function is localized in the $x$ direction
and is spread out in momentum space. 
If the width is large then the wave packet is nearly a plane
wave and the the quantum mechanical interference among spatially 
distant places is important. The width $d_x$
corresponds to the damping factor in the
Kubo formula $\epsilon$ (see Appendix A) in the sense that it causes
decoherence in the system.\cite{finite_kubo2}  
The small damping factor corresponds to small $d_x$. 

In Fig. 7 and Fig. 8,
we show the domain wall width dependence of
the conductance for three values of $\epsilon$ and $d_x$, respectively.
We find that the renormalized conductance is close to one
in the strongly decoherent cases (Fig.7(c) and Fig.8(c)). 
This means that the effect of the domain wall becomes small
as the decoherent effect increases.
Thus we confirm that the intrinsic scattering by the magnetic potential
requires a kind of quantum interference effect. 
Indeed, the increase of the conductivity in the presence of a magnetic
domain wall has been interpreted as an effect of adiabatic motion
(quantum mechanical coherent motion) of the electron
in the magnetic potential.\cite{dw_theo3} 

\section{Effects of the magnetic domains on the conductance}

Making use of the present method, 
let us study the effect of magnetic domain wall on the conductance
in a wire.
The structure of the magnetization profile is given by
\begin{equation}
M_{x}(x) = M_{0} {\rm sech} \left( \frac{x-x_{0}}{\lambda_{w}} \right), 
\label{eq;bloch_wlx}
\end{equation}
\begin{equation}
M_{z}(x) = M_{0} \tanh \left( \frac{x-x_{0}}{\lambda_{w}} \right) \label{eq;blo
ch_wly}.
\end{equation}
Tatara and Fukuyama have calculated analytically the Kubo 
formula for the conductance by a perturbation theory.\cite{dw_theo3} 
They found  
\begin{equation}
\sigma = \frac{e^2}{h} n l \lambda_{\rm{F}} \left[ 1 -  \frac{1}{2 \pi^2}
\frac{{\lambda_{\rm{F}}}^2}{\lambda_w L_x} - \frac{2}{\pi^2} \frac{\lambda_
{\rm{F}}}{l}
\left( \frac{L_w}{L_y} \tan^{-1} \frac{L_x}{\pi L_w} \right) \right] \label{eq;
tatara} ,
\end{equation}
where $L_w \equiv \sqrt{D \tau_w}$,
        $D \equiv \hbar^2 \ k_{\rm{F}}^{2} \tau / 2 m^2$,
        $\tau = l m / \hbar k_{\rm{F}}$, and
        $\tau_w \equiv 24 \pi^2 \lambda_w L_x (\mu_B M_0)^2 \tau / 
(\lambda_{\rm{F}} E_{\rm{F}})^2 $.

We study this problem by the wave packet method
and a numerical evaluation of the
Kubo formula. For this purpose we again use a periodic 
boundary condition and we set a pair of domain walls in the system.

For the numerical calculation we use
a $256 \times 16$ lattice,
corresponding to a physical size of $45 \lambda_{\rm{F}} \times
2.8\lambda_{\rm{F}}$.
Periodic boundary conditions are used.
The impurity concentration is $n_i=4.88 \%$,
$V_0=8.3 E_{\rm{F}}$, $M_0=0.2 \sim 0.4  E_{\rm{F}}$,
the temperature is set to $E_{\rm{F}} / 60$,
and $\epsilon=4.17 \times 10^{-4} E_{\rm{F}}$.
We study eight different impurity configurations and we plot 
the average of the
results. The standard deviation of the data is about $0.1$.
The data strongly depend on the samples. This may be due to the smallness of
the system. 
We find that the numerical data show the enhancement of the conductance,
in agreement with the analytical results
for the single domain wall, as shown in Fig.\ref{fig;kubo_anal_num}.

\section{Summary and Discussion}

We have studied the conductivity of systems with magnetic domain walls
and impurity potential by direct numerical
calculations of the time-dependent Schr\"odinger equation
for the propagation 
of the electron wave packet (wave-packet method).
We found that, when the system has magnetic domain walls, 
the transmission of the wave-packet depends sensitively on the initial 
configuration.
We have proposed a method of a statistical treatment of data
in the spirit of the multi-channel Landauer formula
to obtain the electric conductivity from various initial configurations
of the wave packets.
In order to validate the method, we compare the results with the results
obtained by direct numerical calculations
of the conductivity by the Kubo formula. 
We conclude that the wave-packet method in combination with the statistical 
treatment is a useful approach to obtain information on the conductivity.

We applied the method to a system with a ferromagnetic domain wall, and
found that the analytical estimation of Kubo formula,
numerical calculated Kubo formula and the wave-packet model
consistently show the enhancement of conductivity in the magnetic domain wall.

In the present work, we mainly concentrate on the
comparison between the results of Kubo formula and the wave-packet method.
A numerical evaluation of the Kubo formula requires a lot of computer memory
because we use the simple $L^3$ method, where $L$ is a size of the system.
Memory and CPU time of 
the wave-packet method increase linearly with $L$
and therefore it can be used for 
much large systems and for various boundary conditions.
For instance, preliminary calculations for the three-dimensional case
have been carried out.
The results of the present study show sensitive dependence on the positions of
the impurities.
This may be attributed to the small size of the system whereas
in large lattices we expect self-averaging of the conductivity.

The authors would like to thank Dr. Gen Tatara for valuable discussions.
The present work is partially supported by the Grant-in-Aid from
the Ministry of Education.
The calculation is done at the supercomputer center of ISSP.

\section*{Appendix: Direct calculation of the Kubo formula}
The dc conductivity can be obtained from the Kubo formula,
\begin{eqnarray}
\sigma_{0} &=& \frac{\pi\hbar}{\Omega} \left( \frac{e}{m}\right)
\int \, dE \, \left( - \frac{df(E) }{dE}\right) \sum_{\alpha , \alpha '}
\langle \alpha  | p_{x} | \alpha ' \rangle 
\langle \alpha '| p_{x} | \alpha   \rangle 
\delta \left( E - E_{\alpha}\right) \delta \left( E - E_{\alpha '}\right) ,
\label{kubof}
\end{eqnarray}
where $| \alpha   \rangle $ and $E_{\alpha}$ are the $\alpha$-th 
eigenstate and eigenvalue for the Hamiltonian, and  $f(E)$ is the Fermi
distribution. 

In the Kubo formula approach, 
a non-zero current can be realized by adopting periodic boundary conditions. 
Eigenstates and eigenvalues are obtained
by a direct numerical diagonalization of the system Hamiltonian.

In a numerical approach, we must use a finite system, 
taking finite grids in real space, which leads to discretized 
energy levels.
In this case, it is convenient to introduce the Lorenzian smoothing for
the delta function in (\ref{kubof}):
\begin{eqnarray}
\delta ( x ) \sim \frac{\epsilon}{x^2 + \epsilon^2} ,
\end{eqnarray}
with small $\epsilon$. In case of $\epsilon \rightarrow 0$,
the conductivity vanishes due to the discreteness of the eigenvalues.
In the present study, we take 
$\hbar\epsilon$ to be the same order of energy spacing $\Delta E$:
\begin{eqnarray}
\hbar \epsilon \sim \Delta E .
\end{eqnarray}

\newpage
\begin{figure}
\noindent
\centering
\epsfxsize=7.0cm \epsfysize=6.0cm \epsfbox{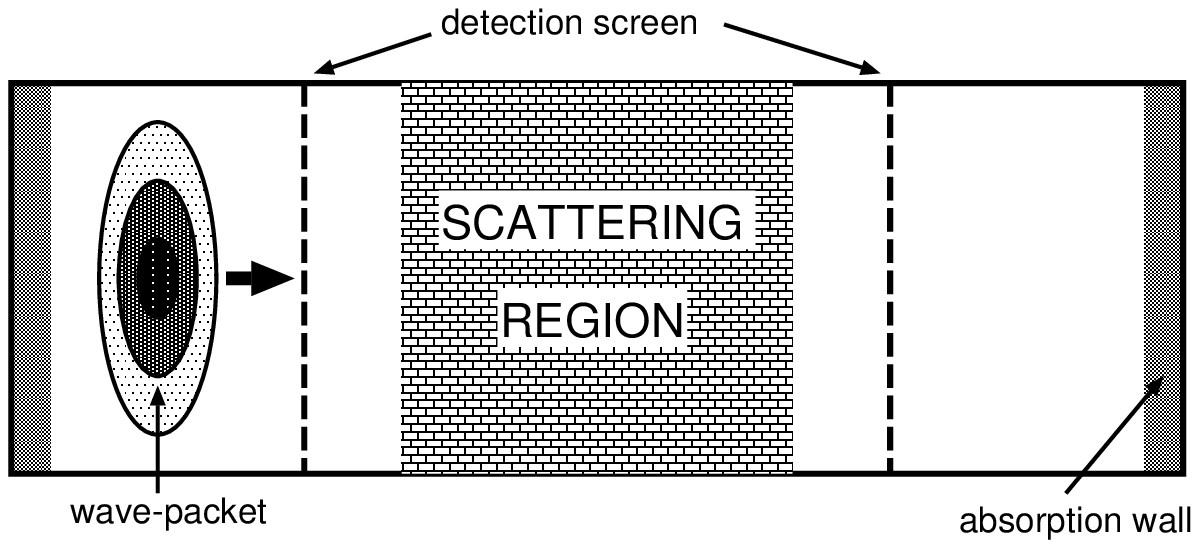} \\
\caption{Simulation model}
\label{config}
\end{figure}

\begin{center}
M. Maeda, K. Saito, S. Miyashita, H. De Raedt
\end{center}

\newpage
\begin{figure}
$$
\begin{array}{ccc}
\epsfxsize=5.0cm \epsfysize=4.0cm \epsfbox{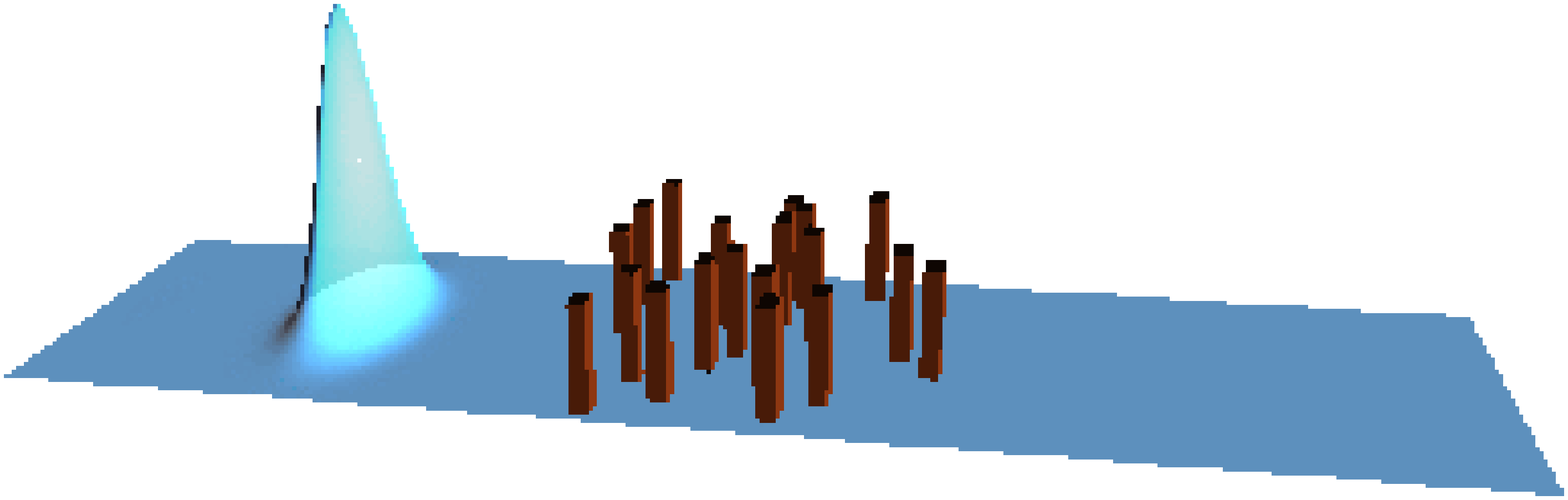}& 
\epsfxsize=5.0cm \epsfysize=4.0cm \epsfbox{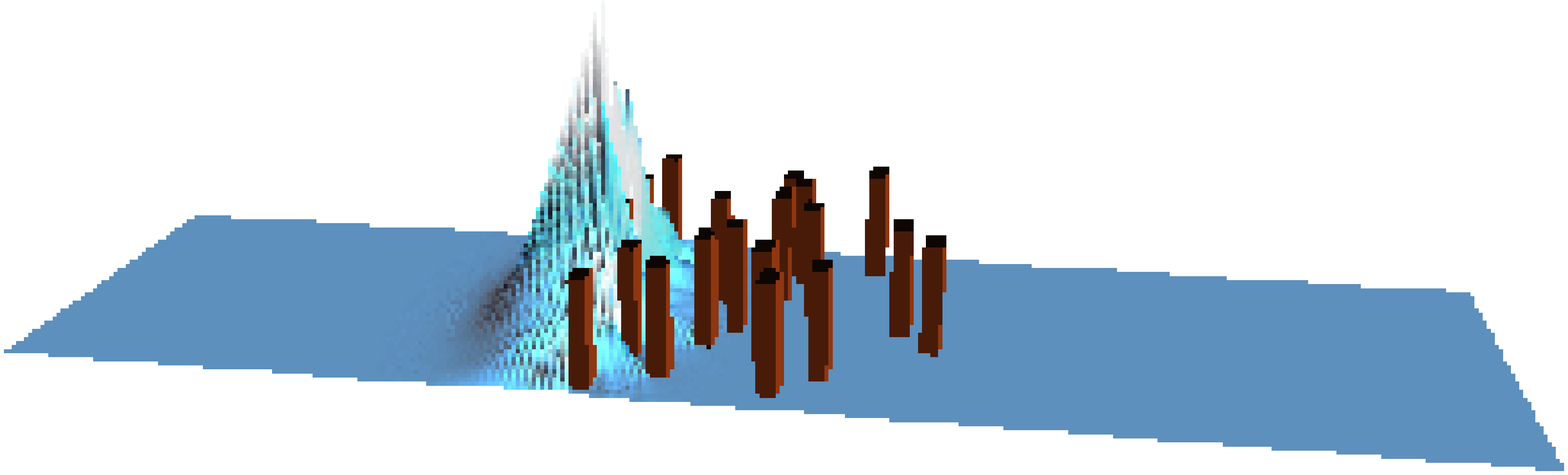}&
\epsfxsize=5.0cm \epsfysize=4.0cm \epsfbox{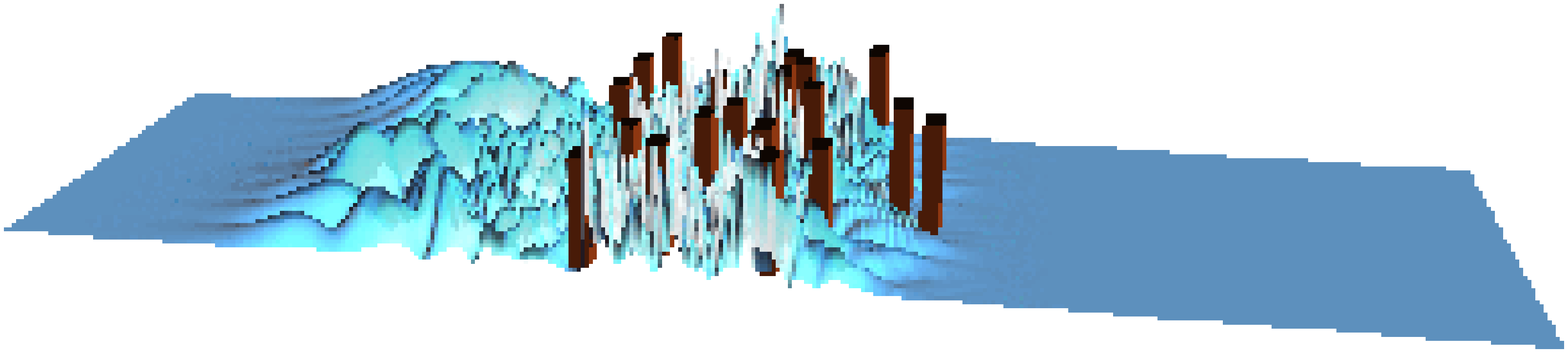}\\ 
{\rm (a)}&{\rm (b)}&{\rm (c)}\\  
\epsfxsize=5.0cm \epsfysize=4.0cm \epsfbox{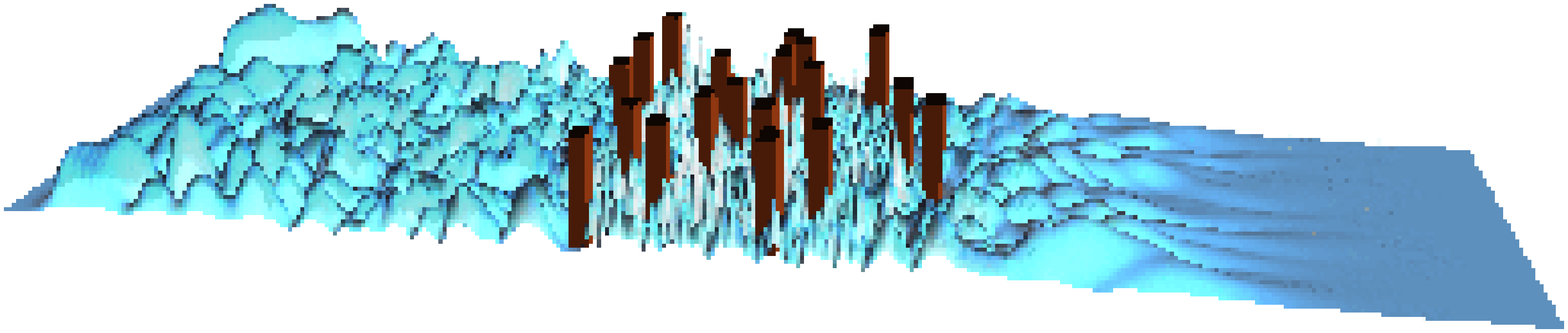}& 
\epsfxsize=5.0cm \epsfysize=4.0cm \epsfbox{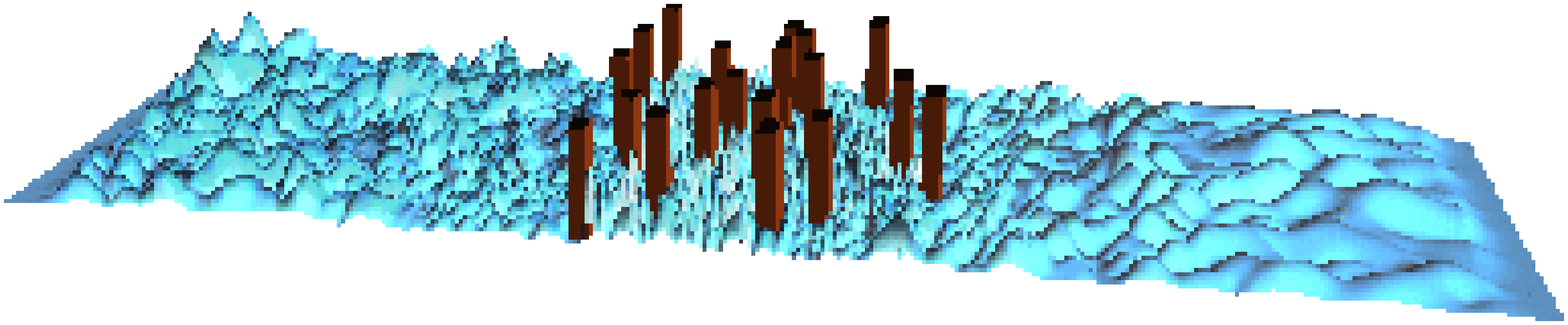}&
\epsfxsize=5.0cm \epsfysize=4.0cm \epsfbox{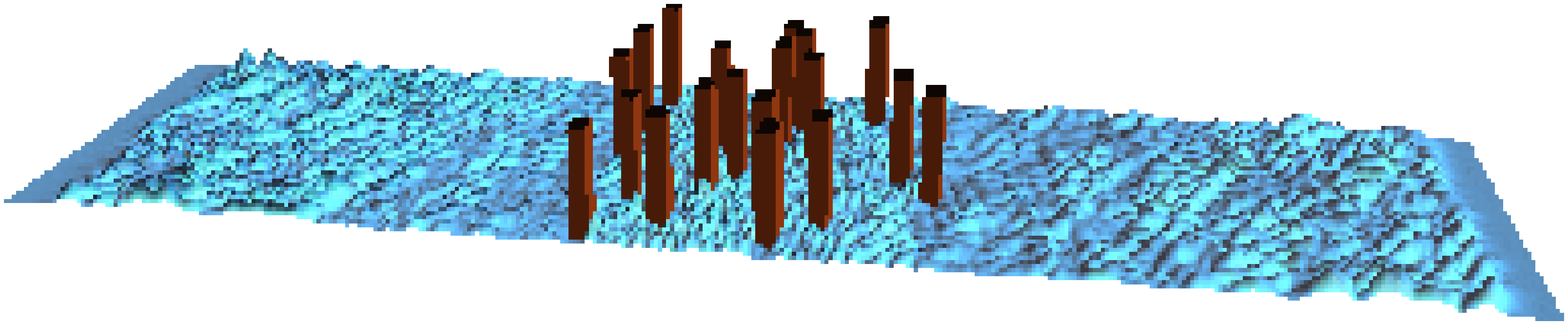}\\ 
{\rm (d)}&{\rm (e)}&{\rm (f)} \end{array}$$
 
\caption{ 2d-picture of the probability distribution: 
(a)$t=10$, (b)$t=55$ ,(c)$t=100$,(d)$t=150$,
(e)$t=200$ and (f) $t=400$.}
\label{2d-picture}
\end{figure}
\begin{center}
M. Maeda, K. Saito, S. Miyashita, H. De Raedt
\end{center}

\newpage
\begin{figure}
\noindent
\centering
\epsfxsize=6.5cm \epsfysize=6.5cm \epsfbox{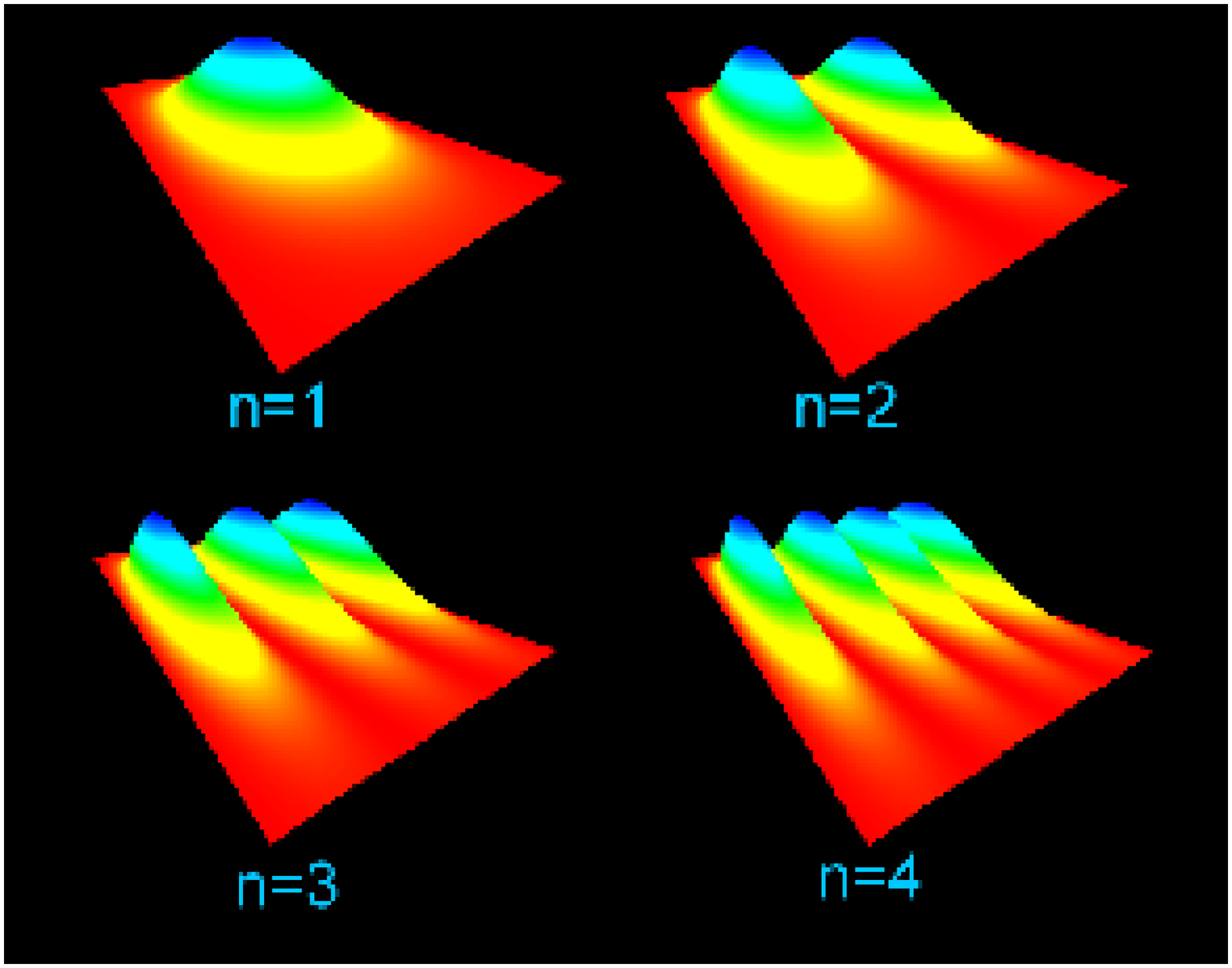} 
\caption{Probability distributions of the initial configurations for
 $k_y=1,2,3$ and 4 in Eq.(\ref{eq:rippled_gaussian})}
\label{initial}
\end{figure}
\begin{center}
M. Maeda, K. Saito, S. Miyashita, H. De Raedt
\end{center}

\newpage 
\quad 
\begin{figure}
\noindent
\centering
\epsfxsize=9.0cm \epsfysize=7.0cm \epsfbox{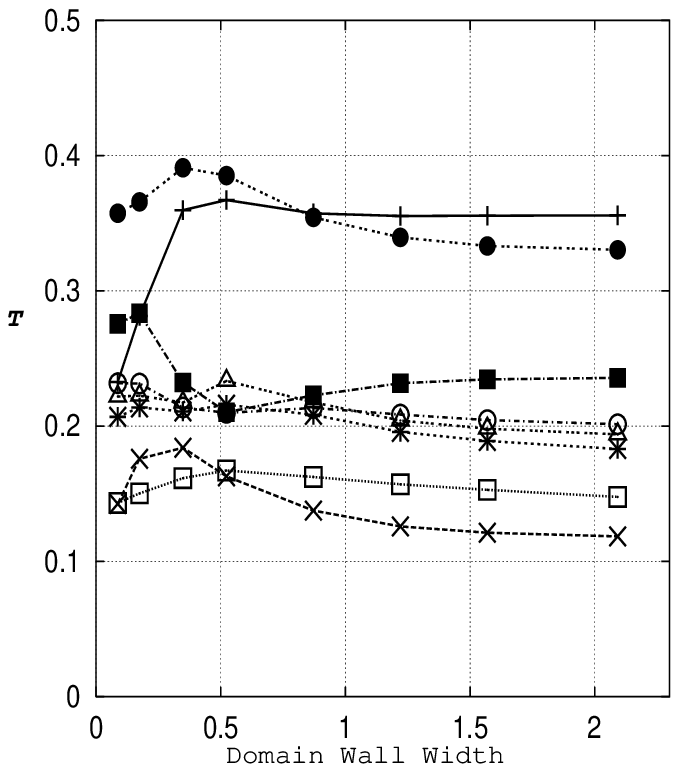} \\
\caption{Dependence on the different initial configuration 
of the transmission coefficient as a function of the domain-wall width.
Each symbol represents the data for different initial condition. 
}
\label{sensitive-dep}
\end{figure}

\begin{center}
M. Maeda, K. Saito, S. Miyashita, H. De Raedt
\end{center}

\newpage 
\begin{figure}[ht]
\noindent
\centering
\epsfxsize=12.0cm \epsfysize=7.0cm \epsfbox{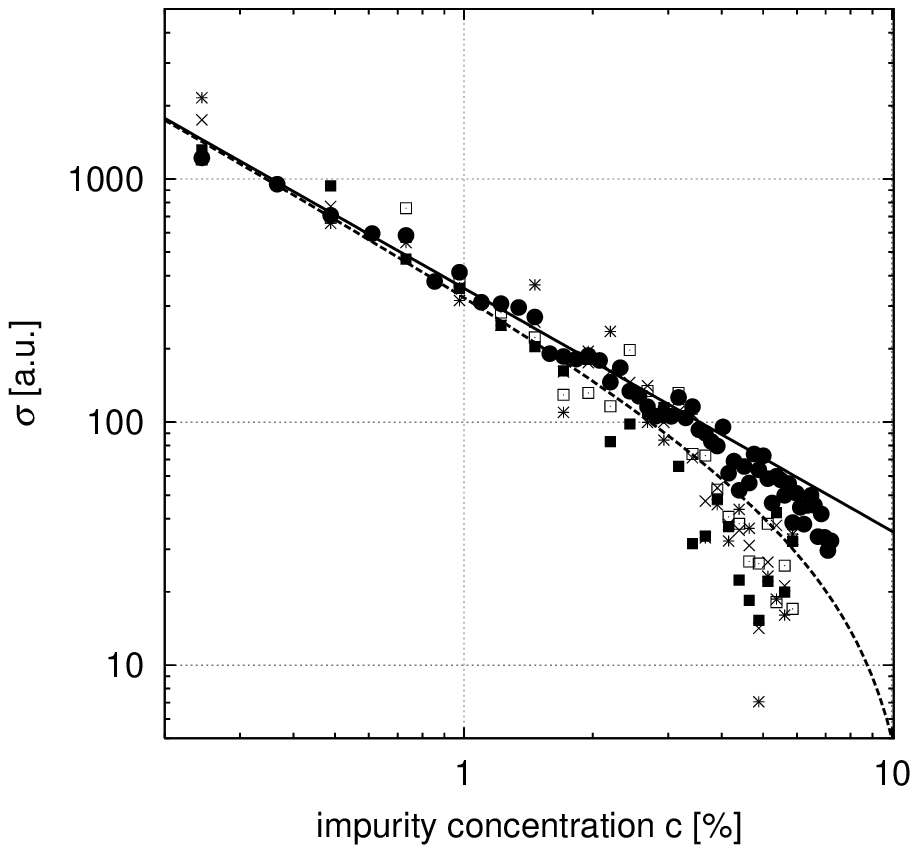}
\caption{Comparison of the transmission coefficients and conductivity.
Kubo formula (solid line), Kubo formula with the
effect of localization (dashed line), Kubo formula numerical method
(closed circle), and the wave packet method with $n=1,2,3$, and $4$
(cross, asterisk, open square, and closed square, respectively).}
\label{fig;kubo_wp_res}
\end{figure}

\begin{center}
M. Maeda, K. Saito, S. Miyashita, H. De Raedt
\end{center}

\newpage 
\begin{figure}
\noindent
\centering
\epsfxsize=13.0cm \epsfysize=6.0cm 
\epsfbox{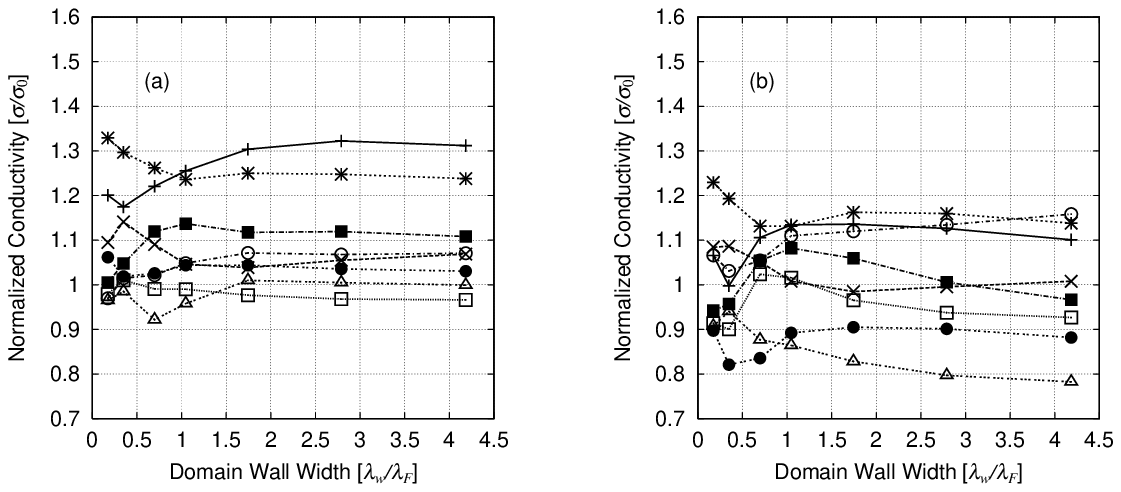}
\caption{Normalized conductance as obtained by the  Kubo formula (a),
and the wave-packet method (b). Different symbols represent
data for different random potentials, using the same symbols in (a) and (b)
for the same random potential.
}
\label{screw}
\end{figure}

\begin{center}
M. Maeda, K. Saito, S. Miyashita, H. De Raedt
\end{center}

\newpage 
\begin{figure}
\noindent
\centering
\epsfxsize=7.0cm \epsfysize=15.0cm \epsfbox{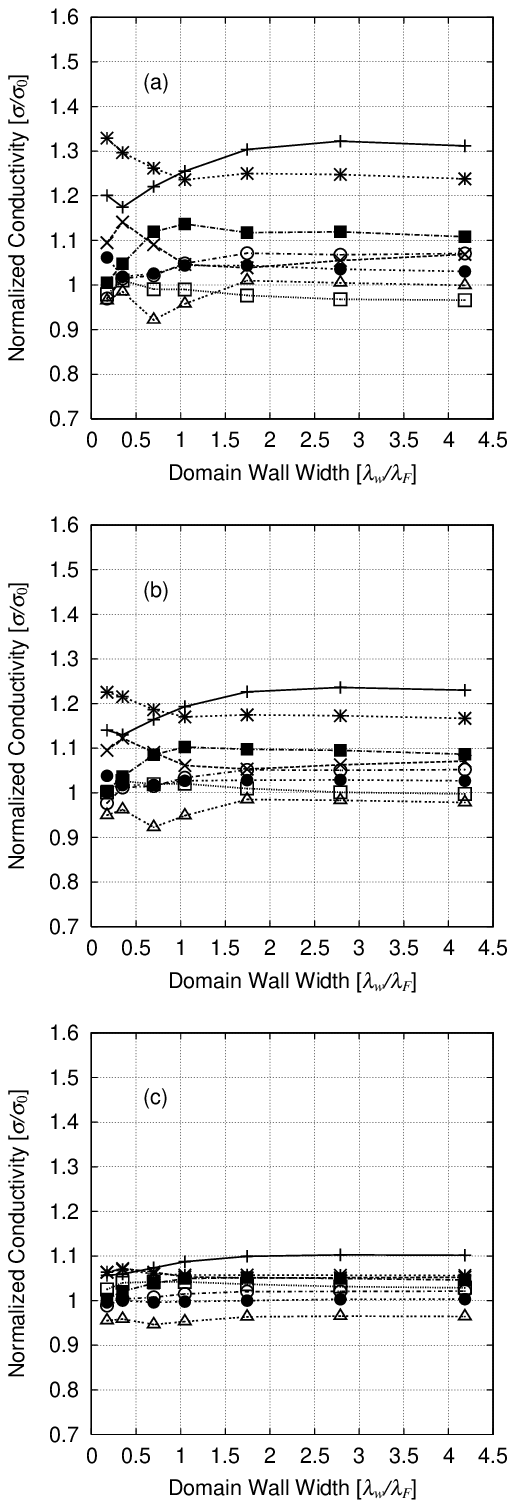}
\label{fig;epsilon}
\caption{Dependence on $\epsilon$ of the normalized 
conductance $\sigma/\sigma_0$. (a):$\epsilon = 4.17\times 10^{-4}$E$_{\rm F}$,(b):
$\epsilon = 8.33\times 10^{-4}$E$_{\rm F}$,(c):$\epsilon = 4.71\times 10^{-3}$E$_{\rm F}$.
Different symbols represent
data for different random potentials, using the same symbols in (a), (b) and (c)
for the same random potential. } 
\end{figure}

\begin{center}
M. Maeda, K. Saito, S. Miyashita, H. De Raedt
\end{center}

\newpage
\begin{figure}
\noindent
\centering
\epsfxsize=7.0cm \epsfysize=15.0cm \epsfbox{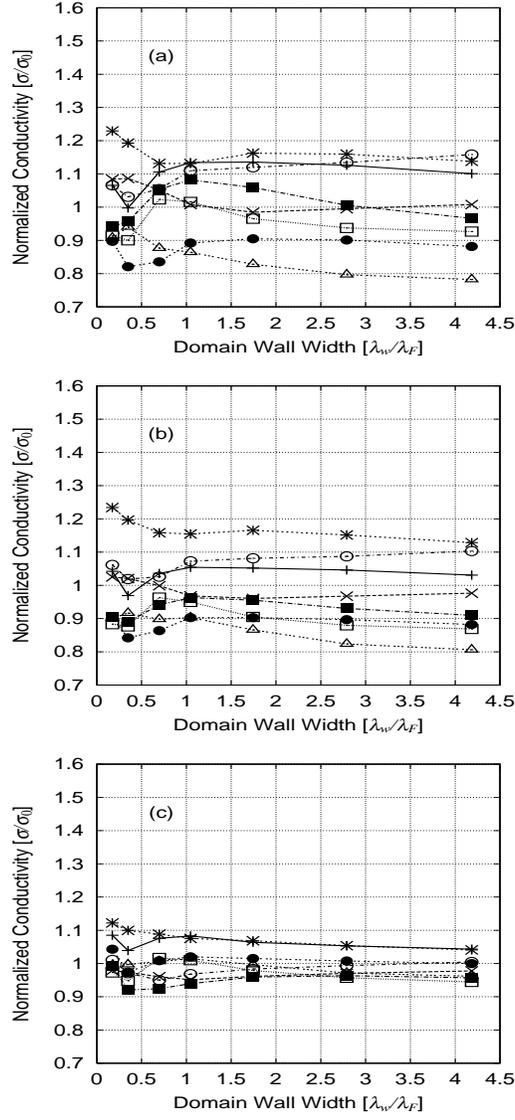}
\label{fig;dx}
\caption{Dependence on $d_{x}$ of the normalized 
conductance $\sigma/\sigma_0$. (a): $d_{x}=3.54 \lambda_{\rm F}$,(b): 
$d_{x}=2.12 \lambda_{\rm F}$,(c):$d_{x}=0.71\lambda_{\rm F}$.
Different symbols represent
data for different random potentials, using the same symbols in (a), (b) and (c)
for the same random potential.} 
\end{figure}

\begin{center}
M. Maeda, K. Saito, S. Miyashita, H. De Raedt
\end{center}

\newpage
\begin{figure}[ht]
\noindent
\centering
\epsfxsize=11.0cm \epsfysize=8.0cm \epsfbox{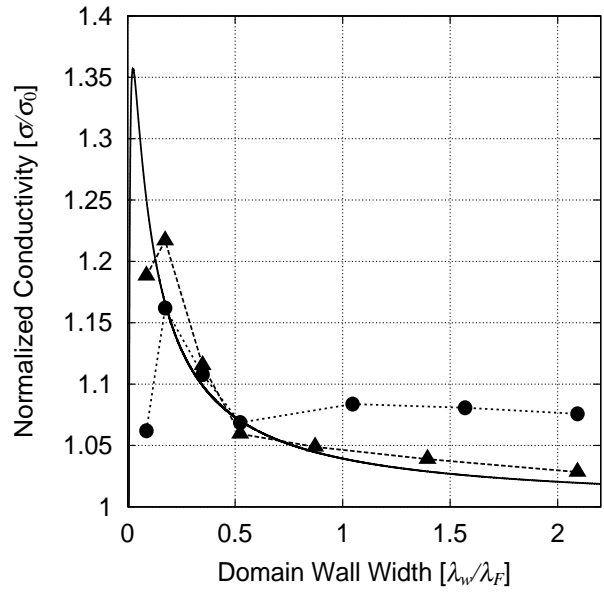}\\
\caption{Comparison between the analytical (solid line) 
and the numerical results of the Kubo formula () and wave packet method().}
\label{fig;kubo_anal_num}
\end{figure}

\begin{center}
M. Maeda, K. Saito, S. Miyashita, H. De Raedt
\end{center}


\begin{thebibliography}{notitle}
\bibitem{GMR-1}M. N. Baibich, J. M. Broto, A. Fert, F. Nguyen Van Dau,
	F. Petroff, P. Eitenne, G. Creuzet, A. Friedrich and
	J. Chazelas: Phys. Rev. Lett. {\bf 61} (1988) 2472.
\bibitem{dw_exp1}
M. Viret, D. Vignoles, D. Cole, J. M. D. Coey, W. Allen, D. S. Daniel
	and J. F. Gregg: Phys. Rev. B {\bf 53} (1996) 8464.

\bibitem{dw_exp2} 
J. F. Gregg, W. Allen, K. Ounadjela, M. Viret, M. Hehn, S. M. Thompson
	and J. M. D. Coey: Phys. Rev. Lett. {\bf 77} (1996) 1580.
\bibitem{dw_exp3}
U. Ruediger and  J. Yu and  S. Zhang and  A. D. Kent and
	S. S. P. Parkin: Phys. Rev. Lett. {\bf 80} (1998) 5639.
\bibitem{dw_exp4}
K. Hong and N. Giordano:
J. Phys. Condens. Matter {\bf 10} (1998) L401.
\bibitem{dw_exp5}
Y. Otani, K. Fukamichi, O. Kitakami, Y. Shimada, B. Pannetier,
	J. P. Nozieres, T. Matuda and A. Tonomura:
Proc. MRS Spring Meeting (San Francisco, 1997) {\bf 475} 215 (1997).
\bibitem{dw_exp6}
Y. Otani, S. G. Kim, K. Fukamochi, O. Kitamich and Y. Shimada:
IEEE Trans. Magn. {\bf 34} (1998) 1096.
\bibitem{dw_theo3}
G. Tatara and H. Fukuyama, Phys. Rev. Lett. {\bf 78} (1997) 3773,
G. Tatara: J. Mod. Phys. B {\bf 15} (2001) 321.
\bibitem{dw_theo12}
P. A. E Jonkers, S. J. Pickering, H. De Raedt and G. Tatara:
Phys. Rev. B {\bf 60} (1999) 15970.
\bibitem{dw_theo4}
P. M. Levy and S. Zhang, Phys. Rev. Lett. {\bf 79} (1997) 5110.
\bibitem{dw_theo5}
A. Brataas and  G. Tatara and G. Bauer, 
Phys. Rev. B {\bf 60} (1999) 3406.
\bibitem{dw_theo6}
J.B.A.N. van Hoof, K. M. Schep, A. Brataas, G. Bauer and P. J. Kelly,
Phys. Rev. B {\bf 59} (1999) 138.
\bibitem{dw_theo7}
G.Tatara, Y.-W.Zhao, M. Mu\~{n}oz and N. Garc\'{\i}a:
Phys. Rev. Lett. {\bf 83} (1999) 2030.
\bibitem{dw_theo8}
P. Grokom and  A. Brataas and G. E. W. Bauer:
Phys. Rev. Lett. {\bf 83} (1999) 4401.
\bibitem{dw_theo9}
H. Imamura and N. Kobayashi and  S. Takasaki and S. Maekawa:
Phys. Rev. Lett. {\bf 84} (2000) 1003.
\bibitem{dw_theo10}
N. Garc\'{\i}a,  M. Mu\~{n}oz and Y. W. Zhao:
Phys. Rev. Lett. {\bf 82} (1999) 2923.
\bibitem{s-t-tdse1} H. De Raedt and K. Michielsen, Computer in Physsics,
{\bf 8} (1994) 600.
\bibitem{s-t-tdse2} H. De Raedt, 
{\it Annual Reviews of Computational Physics IV}, ed. D. Stauffer, 
World Scientific, {\bf 107} (1996).
\bibitem{landauer4} M. Ya. Azbel, J. Phys. C{\bf 14} (1985) L225.
\bibitem{kubo_orig} R. Kubo: J. Phys. Soc. Jpn {\bf 12}(1957) 570.
\bibitem{landauer1} R. Landauer: IBM J. Res. \& Dev. {\bf 1} (1957) 223,
R. Landauer: Z. Phys. B, Condens. Matter {\bf 68} (1987) 217,
M. B\"{u}ttiker, Y. Imry, R. Landauer and S. Pinhas:
Phys. Rev. B {\bf 31} (1985) 6207,
M. Ya. Azbel: J. Phys. C {\bf 14} (1985) L225.
\bibitem{S92}
M. Suzuki, J. Phys. Soc. Jpn. {\bf 61} (1992) 3015.
\bibitem{finite_kubo2}
Y. Imry and N. Shiren: Phys. Rev. B {\bf 33} (1996) 7992.
\end{thebibliography}
\end{document}